# The Role of Hydrogen and Oxygen Interstitial Defects in Crystalline Si cells: Mechanism of Device Degradation in Humid Environment

Bo Li, [a‡] Yu Pang, [a‡] Feifei Zhang, [b] Jinyu Hu, [c] Huiyan Zhao, [b] Guocai Liu, [a] Chao He, *[a] and Xingtao An *[a]

[a.] *School of Science, Hebei Provincial Key Laboratory of Photoelectric Control on Surface and Interface, Hebei University of Science and Technology, Shijiazhuang, 050018, Hebei, China.*

[b.] *Department of Physics and Hebei Advanced Thin Film Laboratory, Hebei Normal University, Shijiazhuang, 050024, Hebei, China.*

[c.] *Key Laboratory of Functional Materials and Devices for Informatics of Anhui Educational Institutions, Department of Physics, Fuyang Normal University, Fuyang 236037, Anhui, China.*

E-mails: hechao210@mails.ucas.ac.cn; anxt2005@163.com.

‡ These authors contributed equally.

**Abstract:** The efficiency of silicon solar cells gradually decreases in various environments, with humidity being a key factor contributing to this decline through moisture-induced degradation (MID) involving multiple mechanisms including encapsulant hydrolysis and metal ion migration. Among these mechanisms, the role of water-derived hydrogen and oxygen interstitial defects represents an underexplored yet fundamental degradation pathway. This study employs density functional theory and quantum transport theory to investigate hydrogen and oxygen interstitial defects as a novel perspective for understanding MID mechanisms. Results reveal that neutral hydrogen interstitials at bond-center sites exhibit low diffusion barriers (0.96 eV) and act as deep-level recombination centers, while oxygen interstitials face higher diffusion barriers (2.2 eV) with limited trapping capability. Device simulations demonstrate that hydrogen defects cause substantially more pronounced photovoltaic current degradation through enhanced non-radiative recombination. Critically, under humid conditions, hydrogen from water molecules readily penetrates silicon lattices forming active recombination centers, while oxygen incorporation remains kinetically limited with negligible impact. This interstitial defect perspective provides novel understanding of MID mechanisms, explaining why moisture exposure primarily degrades silicon solar cells through hydrogen rather than oxygen incorporation, offering fundamental insights for developing targeted mitigation strategies.

# 1 Introduction

Renewable energy technologies have gained increasing prominence as environmentally sustainable alternatives to conventional fossil fuels. Photovoltaic systems represent a critical component of this energy transition, with crystalline silicon (c-Si) solar cells commanding approximately 80% of the global market share[1]. Compared to emerging photovoltaic technologies such as perovskite and other third-generation devices, c-Si solar cells demonstrate superior long-term stability and higher power conversion efficiencies. [2] However, despite their inherent stability advantages, c-Si solar cell performance undergoes gradual degradation over operational lifetimes due to cumulative environmental exposure and intrinsic aging processes.

Comprehensive characterization studies employing accelerated aging protocols, including light soaking and damp heat testing, have been conducted to elucidate the underlying mechanisms responsible for efficiency decline.[3-5] Among the various degradation pathways identified, light-induced degradation (LID),[5] potential-induced degradation (PID),[6] and moisture-induced degradation (MID)[7-9] have emerged as the predominant factors limiting long-term photovoltaic performance under field operating conditions.[10, 11]

MID represents a critical failure mechanism in crystalline silicon photovoltaic modules, where water ingress through module edges and microcracks initiates cascading degradation processes. Ethylene Vinyl Acetate (EVA) encapsulant hydrolysis produces acetic acid and corrosive byproducts that facilitate metal ion migration (Na, Ag, Pb, Sn, Cu, Zn, Al), resulting in formation of oxides, hydroxides, acetates, and complex degradation products[8, 9]. These processes lead to silver grid corrosion, delamination, optical degradation, PID, and increased series resistance[9]. Under field conditions, MID significantly affects module performance and leads to cell degradation. Over time, the prolonged exposure to moisture induces corrosion and oxidation of EVA encapsulants, creating pathways for water ingress[12]. Furthermore, hydrogen atoms derived from moisture can diffuse through the surface $SiO_2$ layer[13] to reach the underlying silicon surface, potentially introducing additional defect states and altering the electronic properties of the semiconductor by migrating into the silicon lattice via vacant crystalline sites. Hydrogen interstitial defects could also arise in c-Si during the crystal growth process[14]. These hydrogen-induced defects may significantly impair battery performance, and their removal from crystalline materials is virtually impossible[14-17]

Oxygen atoms derived from moisture can infiltrate c-Si through the $SiO_2$ on Si interface[18, 19], facilitated by the presence of $OH^-$, $H_2O$, and $H_3O^+$ [20]. This infiltration occurs as oxygen atoms break Si-O bonds in the $SiO_2$ structure and disrupt O-O pairs, leading to the dissociation of these species. Once oxygen enters the $SiO_2$ layer, it typically remains there due to the high energy barrier (approximately 2.5 eV) for further dissociation. Additionally, c-Si inherently contains an interstitial oxygen concentration of approximately $10^{18}$ $cm^{-3}$ [21, 22]. Oxygen's behavior in silicon, including its tendency to occupy fewer interstitial sites [23-25] and form thermal donors[24] under elevated temperatures, significantly influences the electronic properties of silicon, thereby enhancing carrier recombination and contributing to efficiency losses in solar cells.

While extensive investigations have addressed hydrogen and oxygen ingress at silicon interfaces[13, 19, 26] and defect formation energetics[27], a systematic understanding of how moisture-derived H and O interstitial defects collectively impact photovoltaic performance and contribute to MID mechanisms remains elusive. Particularly, the roles of H and O species in governing non-radiative recombination processes through multi-phonon emission pathways have not been comprehensively explored from an electron-phonon coupling perspective. A systematic microscopic analysis linking these interstitial defect configurations to non-radiative transition rates and photovoltaic performance degradation is critically needed to elucidate the fundamental mechanisms underlying MID phenomena. To address this gap, the present study examines the microscopic behavior of hydrogen and oxygen defects in c-Si under humid conditions, systematically assessing their impact on the performance of silicon-based solar cells. This study employs density functional theory (DFT) for electronic structure calculations, climbing image nudged elastic band (CINEB) methods for diffusion pathway analysis, multi phonon emission theory for non-radiative recombination processes, and quantum transport theory to comprehensively explore the behavior of hydrogen and oxygen, as well as their radicals, within the c-Si lattice. The primary objective is to understand how these species contribute to the failure mechanisms of silicon solar cells, particularly in humid environments. By elucidating the microscopic mechanisms of defects in c-Si solar cells, this research provides valuable insights into the degradation of efficiency in practical applications.

# 2 Theoretical Approach

## 2.1 Electronic structure calculations

The structures were optimized using the VASP [28]. The generalized gradient approximation[29] (GGA/PBE) for the exchange-correlation functional was applied to ensure accurate energy calculations and optimized geometries for the systems under investigation. After obtaining the optimized structures, single-point energies were calculated to evaluate the electronic properties at specific configurations. The VASP parameters for the c-Si supercell containing 64 atoms included a plane wave cutoff energy of 450 eV. All calculations employed a 3×3×3 Monkhorst-Pack k-point mesh for Brillouin zone sampling, which converges the total energy to within 2.4 meV/atom. The electronic and ionic convergence criteria were set to $1.56 \times 10^{-7}$ eV/atom and $10^{-5}$ eV/Å, respectively. Calculations were performed in 2×2×2 supercells to minimize finite size effects, and all atoms in the supercell were allowed to relax. All calculations were carried out in reciprocal space. To balance computational efficiency and accuracy, a 2×2×2 supercell was employed in this study, and subsequent validation against corrected 4×4×4 supercell results confirmed that the deviation is negligible, justifying the use of the smaller cell.

Minimum energy paths and diffusion barriers were determined using the climbing-image nudged elastic band (CINEB) method,[30, 31] with three intermediate images. Additional detailed computational parameters and convergence tests are provided in Section S3 of the Supporting Information.

## 2.2 Formation energies

The formation energies of defects were calculated following the method outlined in the literature, identifying the defects that are most readily formed[32]. Additionally, the formation energies[33-35] of defects with different charge states as a function of the Fermi level were also determined.

$$\Delta H_f(\mathrm{X}_i^q)=E(\mathrm{Si_N X};q)-E(\mathrm{Si_N})-n\mu_\mathrm{X}+q\left(E_\mathrm{VBM}-\Delta V_{0/\mathrm{b}}+E_F\right)-E_\mathrm{corr} \quad (1)$$

Here, $E(\mathrm{Si_N})$ represents the total energy of the pristine crystal containing $N$ Si atoms. $E(\mathrm{Si_N X};q)$ denotes the total energy of the crystal containing a defect X with charge state $q$, where X represents either an oxygen or hydrogen interstitial defect. $\mu_X$ is the chemical potential of defect X, and $n$ denotes the number of defects. $E_{VBM}$ represents the energy of the valence band maximum (VBM), while $E_{corr}$ refers to the correction term in the defect formation energy. $\Delta V_{0/b}$ is the potential offset determined by averaging the potential difference in far-field regions.

This highlights the need to carefully consider the limitations and potential errors associated with simulating charged defects in finite-sized supercells, as the primary source of $E_{corr}$ arises from electrostatic interactions between mirror charges. In DFT calculations, finite-sized supercells are typically used to model defects. Due to the finite size of the supercell, charged defects induce electrostatic interactions between the defect and its periodic images, leading to errors in the calculated energies. The Makov-Payne correction method[36, 37] is employed to account for these interactions, specifically $E_{corr}$ , which includes both the leading electrostatic interaction and higher-order correction terms. In 2025, Shang *et al*. further validated the Makov-Payne correction method through calculations on Si, MgO, and $Ga_2O_3$, demonstrating that the Makov-Payne correction method can effectively improve the accuracy of theoretical calculations, making them more consistent with experimental results. Moreover, under certain conditions, it also reduces computational cost.[38]

The correction energy is calculated using a detailed formula provided in formula (S1), while an approximate expression is given in formula (2):

$$E_{\mathrm{corr}} = \frac{q^2\alpha}{2\varepsilon L} \tag{2}$$

where $q$ is the defect charge state, $\alpha$ is the Madelung constant, $\varepsilon$ is the static dielectric constant, and $L$ is the supercell diagonal length. Detailed derivation and higher-order terms are discussed in Section S1 of the Supporting Information. The correction term was computed using equation (2), with the specific values provided in Tables 1 and 2.

To account for finite-size effects in periodic charged defect calculations, we applied the potential alignment method[34, 35] to correct systematic errors in band edge positions. Planar-averaged electrostatic potentials were calculated for both pristine and neutral defect supercells along three crystallographic directions. The potential offset $\Delta V_{0/b}$ was determined by averaging the potential difference in far-field regions, and the corrected valence band maximum $E_{VBM} - \Delta V_{0/b}$ was used in Eq. (1) for defect formation energies. The calculated corrections for all studied configurations are listed in Table 3, with details provided in Section S2 of the Supporting Information.

**Table 1**. The correction values $E_{\mathrm{corr}}$ for the formation energies of hydrogen interstitial defects were calculated using the Makov-Payne correction method. The geometrical definitions of these defect sites are illustrated in Fig. 1 and described in detail in Section 3.1.

| H SITES | Corrected energy for different H charge states (eV) | |
|---|---|---|
| | +1 | -1 |
| T[a] | 0.02 | 0.02 |
| BC[b] | 0.02 | 0.02 |
| C[c] | 0.02 | 0.02 |
| M[d] | 0.02 | 0.02 |
| H[e] | 0.02 | 0.02 |
| AB[f] | 0.02 | 0.02 |

Notes:
*a* Tetrahedral site, located at the center of a tetrahedral void coordinated by four Si atoms.
*b* Bond-centered site at the midpoint of a Si–Si bond.
*c* Cage-centered site, located at the center of the line connecting adjacent M sites within the cage region.
*d* Midpoint site, located at the midpoint between a tetrahedral site and its nearest Si atom.
*e* Hexagonal site, located at the center of a hexagonal face formed by surrounding Si atoms.
*f* Antibonding site, located along the ⟨111⟩ direction opposite to a Si–Si bond.

**Table 2**. The correction values $E_{\mathrm{corr}}$ for the formation energies of oxygen interstitial defects were calculated using the Makov-Payne correction method. Here, BC1 denotes the midpoint (bond-center) site, which is located at the midpoint of a Si–Si bond.

| SITES | Corrected energy for different O charge state (eV) | | | |
|---|---|---|---|---|
| | +2 | +1 | -1 | -2 |
| BC1 | 0.05 | 0.02 | 0.02 | 0.05 |

**Table 3.** Potential alignment corrections ($\Delta V_{0/\mathrm{b}}$) for neutral hydrogen and oxygen interstitial defects at various high-symmetry sites in the $2 \times 2 \times 2$ silicon supercell. The corrections represent the electrostatic potential difference between the defective and pristine bulk regions, applied to align band edges in defect formation energy calculations.

| Corrected term (eV) | Sites for H and O interstitial defects | | | | | | |
|---|---|---|---|---|---|---|---|
| | $\mathrm{H}^0_{i(\mathrm{AB})}$ | $\mathrm{H}^0_{i(\mathrm{BC})}$ | $\mathrm{H}^0_{i(\mathrm{C})}$ | $\mathrm{H}^0_{i(\mathrm{H})}$ | $\mathrm{H}^0_{i(\mathrm{M})}$ | $\mathrm{H}^0_{i(\mathrm{T})}$ | $\mathrm{O}^0_{i(\mathrm{BC1})}$ |
| $\Delta V_{0/\mathrm{b}}$ | -0.003 | -0.014 | 0.014 | 0.011 | -0.004 | -0.002 | -0.025 |

## 2.3 Chemical potential

Following the determination of defect formation energies and associated correction terms, we construct and analyze chemical potential phase diagrams to assess the thermodynamic stability of defects. The atomic chemical potential is defined as[32]:

$$\mu_{\mathrm{X}} = E_{\mathrm{X}} + \Delta\mu_{\mathrm{X}} \tag{3}$$

Here, $E_{\mathrm{X}}$ represents the reference energy per atom of X in its natural state. $\Delta\mu_{\mathrm{X}}$ represents the deviation from this reference and reflects the growth conditions. To prevent elemental precipitation, $\Delta\mu_{\mathrm{X}} \leq 0$, and its value is further limited by the thermodynamic stability of the host compound. For hydrogen, $E_{\mathrm{H}}$ is the energy per atom in $H_2$, and for oxygen, $E_O$ is the energy per atom in $O_2$.

The following secondary items should not be considered for precipitation: $SiH_4$, $Si_2H_6$, $H_2$, SiO, $SiO_2$, $SiO_4$, $O_2$, as they do not form stable defect structures in crystalline silicon. Additionally, since the study focuses on the behavior of individual interstitial defects within the crystal, it is crucial to establish the chemical potential ranges, which can be determined using the following calculation formula.

$$\Delta\mu_{Si}+4\Delta\mu_{H}<\Delta H(SiH_4)=-9.41eV \quad (4)$$

$$2\Delta\mu_{Si}+6\Delta\mu_{H}<\Delta H(Si_2H_6)=-14.13eV \quad (5)$$

$$\Delta\mu_{H} \leq 0 \quad (6)$$

$$\Delta\mu_{Si} + \Delta\mu_{O} < \Delta H(SiO) = -1.28eV \quad (7)$$

$$\Delta\mu_{Si} + 2\Delta\mu_{O} < \Delta H(SiO_2) = -8.43eV \quad (8)$$

$$\Delta\mu_{Si} + 4\Delta\mu_{O} < \Delta H(SiO_4) = -1.92eV \quad (9)$$

$$\Delta\mu_{O} \leq 0 \quad (10)$$

$$\Delta\mu_{Si} > 0 \quad (11)$$

## 2.4 Defect Concentration

Defects within a crystal tend to reach a thermodynamic equilibrium, leading to a stable presence of defects characterized by their concentration. Due to variations in the thermodynamic stability of different types of defects, their equilibrium concentrations within the crystal can differ significantly. Since defect concentration is determined by not only the intrinsic properties of the defects themselves, but also varies with temperature. The calculation of defect concentration is given by Eq. (12)[39, 40]:

$$N_D=g\cdot N_S\cdot e^{-\frac{\Delta H_f}{k_B T}} \quad (12)$$

Here, $g$ denotes the total degeneracy factor accounting for the orientational and spin multiplicity of the defect, $k_B$ is the Boltzmann constant, $T$ is the thermodynamic temperature, and $\Delta H_f$ represents the defect formation energy. $N_S$ represents the concentration of available sites per unit volume that can be occupied by the defect. For interstitial defects, $N_S$ corresponds to the number of symmetry-inequivalent interstitial positions in the perfect crystal lattice that can accommodate the interstitial atom. These sites are determined by the crystal structure and are obtained by enumerating all crystallographically distinct interstitial positions within the primitive unit cell.

The total degeneracy $g$ reflects the symmetry characteristics of the system and is defined as shown in Eq. (13):

$$g=g_{sites}\times g_{orient}\times g_{irrep}\times g_{spin} \quad (13)$$

Here, $g_{sites}$ refers to the site degeneracy, which reflects the number of all symmetrically equivalent sites; $g_{origent}$ denotes the orientational degeneracy, representing the number of equivalent configurations in different spatial orientations that possess the same formation energy and physical properties; $g_{irrep}$ corresponds to the irreducible representation degeneracy, which arises from group-theoretical classification of equivalent states; and $g_{spin}$ refers to the spin degeneracy, indicating the number of equivalent spin orientations available for a given defect state.

## 2.5 Defect Transition Levels

The charge state of a defect is not fixed and can change depending on the position of the Fermi level. This change in charge state at a specific Fermi level is referred to as the defect transition level[41-43]. The calculation of the defect transition level is given by Eq. (14).

$$\varepsilon(q_1/q_2) = \frac{\Delta H_f(X^{q1};E_F=0)-\Delta H_f(X^{q2};E_F=0)}{q_2-q_1} \tag{14}$$

Here, $q_1$ denotes the initial charge state of the defect, while $q_2$ represents the final charge state after the transition. $\Delta H_f(X^q;E_F=0)$ refers to the formation energy of defect $X$ in charge state $q$ when the Fermi level is at the valence band maximum. The position of the defect transition level can be used to evaluate whether a defect is detrimental. A transition level $\varepsilon(q_1/q_2)$ located near the middle of the band gap indicates favorable thermodynamic conditions for charge state transitions in this energy range. The actual recombination behavior, however, is determined by the positions of the defect states and their capture cross-sections, which together influence carrier lifetimes and optoelectronic performance.

## 2.6 Photocurrent

To provide a clearer understanding of the impact of interstitial defects at different sites on the photocurrent of Si-based solar cells, a device was constructed using (100) Si with and without interstitial hydrogen or oxygen defects. The device consists of a left lead (L), a central region (CR), and a right lead (R), as shown in Fig. S5. Subsequently, the Atomistix ToolKit (ATK) software[44], based on DFT combined with the non-equilibrium Green's function (NEGF) method[45], was used to compute the photovoltaic current. This method allows for the simulation of electronic transport properties under non-equilibrium conditions. The combination of VASP for structural optimization and ATK for transport properties enables a comprehensive analysis of the photovoltaic performance at the atomic level.

All DFT+NEGF calculations in this study were performed using the ATK software, which used the GGA+1/2 exchange-correlation method[46] and PBE functional[47]. The photocurrent was then calculated as a first-order perturbation to the device's electronic system by introducing the electron-photon interaction Hamiltonian[48]:

$$H' = \frac{e}{m_0}\mathbf{A}\cdot\mathbf{P} \tag{15}$$

where $\mathbf{A}$ is the vector potential for a monochromatic light and $\mathbf{P}$ is the momentum operator.

$$\mathbf{A} = \mathbf{e}\left(\frac{h\sqrt{\tilde{\mu}_r\tilde{\varepsilon}_r}}{4\pi N\omega c\tilde{\varepsilon}}F\right)^{\frac{1}{2}}\left(be^{i\omega t} + b'e^{-i\omega t}\right) \tag{16}$$

where $\tilde{\mu}_r$ is relative permeability, $\tilde{\varepsilon}_r$ and $\tilde{\varepsilon}$ is the relative permittivity and permittivity, $\omega$ is the light frequency, $F$ is the photon flux, $N$ is the number of photons, $b$ and $b'$ represent the bosonic annihilation and creation operators, respectively. $\mathbf{e}$ is the direction of $\mathbf{A}$, which is determined by the polarization of the field. In this work, $\tilde{\varepsilon}_r = 13.1$ was adopted, based on first-principles calculations, and a photon flux $F$ of $1\mathrm{s}^{-1}\text{Å}^{-2}$. The polar light energy ranged from $0$ to $5.5\mathrm{eV}$.

The perturbation Hamiltonian $H'$ can be added to the central region Hamiltonian $H_{\mathrm{CR}}$ for the next DFT+NEGF calculations using open boundary conditions along the photocurrent direction. The Hamiltonian $H_{\mathrm{L}}$ for left lead and $H_{\mathrm{R}}$ for right lead can be computed using periodic boundary conditions within the DFT framework. In the entire device calculations above, the k point's mesh was set as $3\times3\times87$ to ensure the energy convergence to be $1\times10^{-5}$eV/atom. Finally, the self-consistent spectral density matrix of the central region can be obtained:

$$\rho(E)^{\mathrm{L(R)}} = \frac{1}{2\pi} G(E)\Gamma(E)^{\mathrm{L(R)}} G^{\dagger}(E) \tag{17}$$

It includes electronic density contributions from both the left and right leads. The retarded Green's function matrix, $G(E)$, incorporates the effects of the lead states on the central region, referred to as the self-energy[49] $\Sigma(E)^{\mathrm{L(R)}}$:

$$G(E) = [(E + i\delta_{+})S - H_{\mathrm{CR}} - \Sigma(E)^{\mathrm{L}} - \Sigma(E)^{\mathrm{R}}]^{-1} \tag{18}$$

where $\delta_{+}$ is an infinitesimal positive number, and $S$ represents the overlap matrix of the central region. The broadening function of the left (right) lead is given by:

$$\Gamma(E)^{\mathrm{L(R)}} = i\left[\Sigma(E)^{L(R)} - \Sigma(E)^{L(R)^{\dagger}}\right] \tag{19}$$

Then, the current into the left and right leads due to the absorption of photons is given by[50, 51]:

$$I_{\alpha} = \frac{e}{h}\int_{-\infty}^{+\infty} \sum_{\alpha,\beta=\mathrm{L,R}} [1 - f_{\alpha}(E)]\, f_{\beta}(E - \hbar\omega) T_{\alpha,\beta}^{-}(E) - f_{\alpha}(E)\left[1 - f_{\beta}(E + \hbar\omega)\right] T_{\alpha,\beta}^{+}(E)\mathrm{d}E \tag{20}$$

where $f_{\alpha}$ and $f_{\beta}$ are the Fermi distribution functions of left and right leads. The transmission coefficient $T_{(\alpha,\beta)}(E)$ is given by[52]:

$$T_{(\alpha,\beta)}(E) = \mathrm{Tr}[\Gamma(E)^{L} G(E) \Gamma(E)^{R} G(E)^{\dagger}] \tag{21}$$

## 2.7 Non-radiative transitions

Non-radiative carrier recombination involves electronic transitions between defect states through multi phonon emission, where excess energy is dissipated as lattice vibrations. This process governs carrier capture coefficients, capture cross sections, and recombination rates in defective semiconductors. The theoretical framework and computational implementation are realized through the Nonrad code[53], which provides first-principles calculations of non-radiative recombination processes in crystalline materials.

The non-radiative capture coefficient is described by Fermi's golden rule[53, 54]:

$$C = f\frac{2\pi}{\hbar} g V W_{\mathrm{if}}^{2} \sum_{m} w_{m} \sum_{n} \left|\langle \chi_{im} | \hat{Q} - Q_{0} | \chi_{fn} \rangle\right|^{2} \times \delta(\Delta E + m\hbar\Omega_{\mathrm{i}} - m\hbar\Omega_{\mathrm{f}}) \tag{22}$$

where $g$ is the degeneracy factor, $V$ is the supercell volume, and $w_{\mathrm{m}}$ denotes thermal occupation, $\langle \chi_{im} | \hat{Q} - Q_{0} | \chi_{fn} \rangle$ are phonon matrix elements. The delta function ensures energy conservation during the transition.

The multidimensional phonon problem is reduced to a one-dimensional configuration coordinate model using an effective coordinate $Q$ that captures the dominant lattice distortion: $Q = \sqrt{\sum_{\alpha} m_{\alpha} \Delta R_{\alpha}^{2}}$, where the sum is over atoms $\alpha$ with mass $m_{\alpha}$ and a displacement from equilibrium of $\Delta R_{\alpha}$. It is defined as the mass-weighted structural distortion between the relaxed geometries of the initial and final charge states[53, 54]. This effective mode naturally captures the collective contributions from all phonon modes coupled to the electronic transition. The electron-phonon coupling matrix element is calculated within density functional theory: $W_{if} = \langle \psi_{i} | \partial_{Q} \hat{H} | \psi_{f} \rangle$. The capture cross section is obtained by $\sigma =$ $C\ /\ \bar{v}$, where the average thermal velocity $\bar{v}$ is evaluated using the Nonrad code.

The calculated capture coefficients enter the Shockley-Read-Hall formalism[55]:

$$R_{\mathrm{SRH}} = \frac{np - n_i^2}{(n+n_i)/C_p N_D + (p+p_i)/C_n N_D} \tag{23}$$

where $C_{\mathrm{n,p}}$ are the electron and hole capture coefficients, and $N_{\mathrm{D}}$ is the defect concentration and other parameters follow the definitions given by Yuan *et al*[55].

# 3 Results and Discussion

## 3.1 Impact of H interstitial defects

Hydrogen interstitial behavior in crystalline silicon remained poorly understood until the comprehensive theoretical analysis by Estreicher *et al*. in 1994[56]. This landmark study identified six energetically stable interstitial sites for hydrogen within the silicon lattice, predominantly located at high-symmetry crystallographic positions: (i) The tetrahedral interstitial site (T), positioned at the center of the tetrahedral void formed by four nearest-neighbor Si atoms [Fig. 1(a)] [57]; (ii) the bond-center site (BC), located at the midpoint of Si-Si bonds [Fig. 1(b)] [58]; (iii) the anti-bonding site (AB), situated along the <111> direction opposite to Si-Si bonds, displaced 0.42 Å from the T position [Fig. 1(c)] [59]; (iv) the hexagonal site (H), positioned at the center of hexagonal faces formed by silicon atoms [Fig. 1(d)] [60]; (v) the mid-site (M), located at the midpoint between T sites and nearest Si atoms [Fig. 1(e)] [61]; and (vi) the C-site, positioned at the center of lines connecting adjacent M sites [Fig. 1(f)] [61].

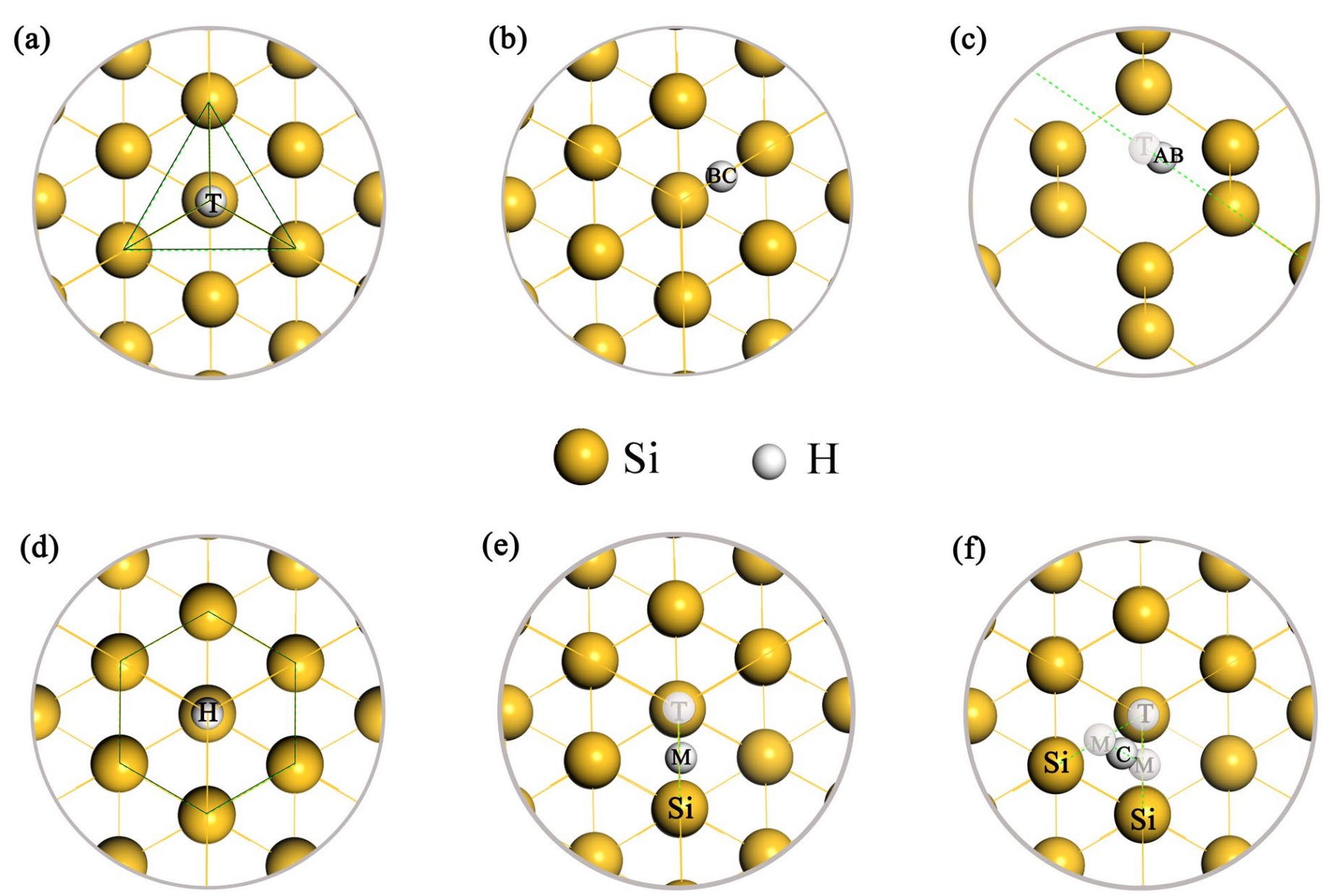

**Figure 1**. Interstitial hydrogen located at six high symmetry sites : (a) T (the center of the space formed by the tetrahedron of the nearest four of Si atoms)、(b) BC (Bond-Center)、(c) AB (Anti-bonding)、(d) H (Hexagonal)、(e) M (Mid) and (f) C (Connecting) within the c-Si.

Based on these established interstitial sites, hydrogen atoms were placed at each of the six positions described above and subjected to structural relaxation and self-consistent field calculations in VASP for three charge states (-1, 0, and +1). The relaxation calculations revealed that $H_i^{+1}$ initially positioned at both the T and C sites spontaneously migrated to the bond-center (BC) configuration during geometry optimization, as shown in Fig. S6. Therefore, $H_{i(T)}^{+1}$ and $H_{i(C)}^{+1}$ configurations can be effectively considered as $H_{i(BC)}^{+1}$ in subsequent analysis.

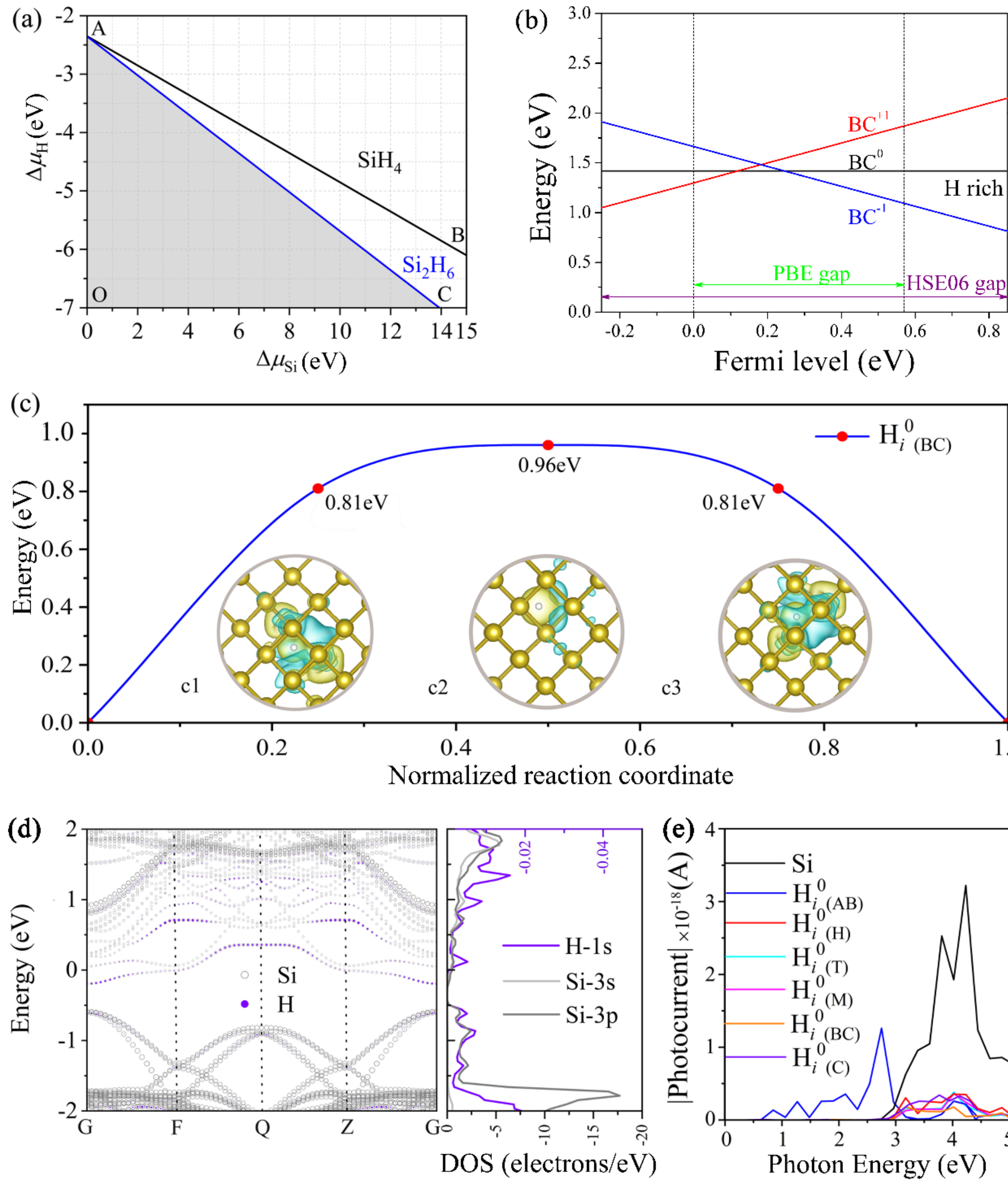

**Figure 2**. The chemical potential range for (a) $\Delta\mu_{Si}$ and $\Delta\mu_H$ doping in the c-Si; (b) the formation energies of $H_{i(BC)}^{-1}$, $H_{i(BC)}^{0}$ and $H_{i(BC)}^{+1}$; (c) the diffusion barrier of the pathway for hydrogen from the BC site to the nearest BC site, along with the corresponding differential charge density for the diffusion path; and (d) the combination of projected bands and projected density of states for $H_{i(BC)}^{0}$;

(e) photocurrent intensity figure of neutral hydrogen dopant defects at different sites with a device cross-sectional area of $2.91 \times 10^{-18}\ \mathrm{m}^2$.

**Formation Energy Analysis for Hydrogen Defects**. Prior to defect formation energy calculations, we established the electronic structure parameters and chemical potential references for the host silicon matrix. Band structure calculations using HSE06 functional yielded a band gap of 1.11 eV (Fig. S7(b)), more accurate than the PBE value of 0.57 eV and consistent with literature[62]. This HSE06 band gap defines the accessible Fermi level range for subsequent defect formation energy analysis.

To establish appropriate chemical potential constraints for hydrogen defect formation energy calculations, we constructed the stability diagram shown in Fig. 2(a). The black and blue lines represent the stability boundaries for $SiH_4$ and $Si_2H_6$ phases, respectively, with the allowed thermodynamic region (AOCA) indicating conditions where no competing secondary phases form. Under these constraints, the chemical potential range is determined as $-\infty < \Delta\mu_{\mathrm{H}} < -2.35$ eV, corresponding to hydrogen-rich conditions with $\mu_{\mathrm{H}} = -3.51$ eV (Eq. 3). These chemical potential values were then applied to Eq. (1) to calculate defect formation energies for different charge states (-1, 0, +1) at various interstitial sites as a function of Fermi level.

Having established the chemical potential constraints, we validated our computational approach by evaluating finite-size effects in defect formation energy calculations. Benchmark calculations of $\mathrm{H}_{i(\mathrm{BC})}$ defects using 2×2×2 and 4×4×4 supercells with MP electrostatic corrections show formation energy deviations of only 0.02 eV ($\mathrm{H}^{0}_{i(\mathrm{BC})}$) and 0.12 eV ($\mathrm{H}^{+1}_{i(\mathrm{BC})}$) (Fig. S8), confirming the reliability of our methodology for both neutral and charged defect calculations.

With the validated computational methodology, we systematically examined hydrogen interstitial formation energies across different sites to identify the most thermodynamically favorable configurations. Formation energy calculations were performed for T, BC, C, AB, M, and H interstitial sites, with T and C sites found to be equivalent to the BC configuration. As shown in Fig. S9, comprehensive analysis reveals that BC and AB sites exhibit superior thermodynamic stability compared to other positions, attributed to the presence of charge state transition levels within the silicon band gap. These transitions enable Fermi level pinning effects[63, 64], providing stabilization mechanisms absent in T, C, M, and H sites which maintain fixed charge states. Note that hydrogen diffusion pathways between H sites along <111> directions were excluded from analysis due to prohibitively long migration distances under practical conditions[60].

For the BC site, $\mathrm{H}^{0}_{i(\mathrm{BC})}$ emerges as the most stable configuration with a formation energy of 1.42 eV[65] (Fig. 2(b) and Fig. S9(b)). This value represents the lowest formation energy among all neutral hydrogen configurations across different interstitial sites examined in this work, consistent with previous theoretical predictions identifying the BC position as the global energy minimum for $\mathrm{H}^{0}_{i(\mathrm{BC})}$[66]. The formation energy diagram reveals distinct stability regions: $\mathrm{H}^{+1}_{i(\mathrm{BC})}$ dominates for Fermi levels from -0.25 to 0.12 eV, $\mathrm{H}^{0}_{i(\mathrm{BC})}$ becomes favorable between 0.12 and 0.24 eV, and $\mathrm{H}^{-1}_{i(\mathrm{BC})}$ is most stable from 0.24 to 0.82 eV, with charge transition levels at 0.12 eV (+1/0) and 0.24 eV (0/-1) calculated using Eq. (14). The Fermi level pinning between 0.12 and 0.24 eV indicates that $\mathrm{H}^{0}_{i(\mathrm{BC})}$ is the most thermodynamically favorable state under equilibrium conditions. The calculated charge transition levels at the BC site are illustrated in Fig. S10(b).

Similarly, for the AB site, $\mathrm{H}^{+1}_{i(\mathrm{AB})}$ shows a relatively low formation energy of 1.48 eV and undergoes charge transition to $H_{i(AB)}^{-1}$ at 0.16 eV. Given that the Fermi level is typically constrained within the 0 to 0.12 eV range, $H_{i(AB)}^{+1}$ represents the most stable configuration at the AB site.

**Hydrogen Diffusion Kinetics and Migration Barriers.** To complement the thermodynamic analysis, hydrogen diffusion kinetics were investigated using CINEB calculations to determine migration barriers between interstitial sites. Before performing the CINEB calculations, we conducted convergence tests for the initial and final diffusion sites and the number of selected images (Fig. S11).

For the neutral charge state, $H_{i(BC)}^{0}$ has a formation energy of 1.42 eV and exhibits a significantly lower diffusion barrier of 0.96 eV (Fig. 2(c)). The calculated barrier is 0.12 eV higher than the value of 0.84 eV reported by Dewar *et al*[14, 67] using the MINDO/3 method[68], but remains in reasonable agreement considering the difference in computational methods. This discrepancy can be attributed to methodological differences: MINDO/3, being a semi-empirical approach, approximates electronic interactions and may underestimate barrier heights, whereas our CINEB calculations employ first-principles methods that explicitly account for energy variations along the entire diffusion pathway. The differential charge density analysis reveals significant electron redistribution during migration, with electron depletion around hydrogen (blue regions) and accumulation near silicon atoms (yellow regions) in Fig. 2(c1-c3), demonstrating that Si-H orbital interactions critically determine the calculated barrier height. The BC-to-BC diffusion pathway (0.96 eV barrier) represents the most favorable route for hydrogen migration in c-Si, consistent with the BC-C(or M)-BC mechanism reported by Walle *et al.* [69], confirming that BC site connectivity governs hydrogen transport in silicon.

The migration barriers for $H_{i(BC)}^{+1}$ (Fig. S12(b)) and $H_{i(BC)}^{-1}$ (Fig. S13(a)) are 1.06 eV and 1.00 eV, respectively, comparable to the 0.96 eV barrier for $H_{i(BC)}^{0}$ (Fig. 2(c)). The narrow range (0.96–1.06 eV) indicates minimal charge-state dependence for diffusion kinetics at the BC site.

While the neutral state exhibits a higher formation energy (1.42 eV) than the charged states, $H_{i(BC)}^{0}$ becomes the dominant species in intrinsic silicon due to Fermi level pinning. The charge transition levels $\varepsilon(+/0)$ and $\varepsilon(0/-)$ of interstitial hydrogen lie deep within the band gap, placing the mid-gap Fermi level within the stability range of the neutral charge state[70, 71]. This pinning effect thermodynamically stabilizes $H_{i(BC)}^{0}$ despite its higher absolute formation energy. Combined with the lowest migration barrier (0.96 eV), neutral hydrogen therefore dominates hydrogen-related processes in near-intrinsic silicon, including diffusion, defect passivation, and hydrogen-vacancy complex formation.

At the AB site, the formation energies of charged hydrogen states cross at $E_f = 0.12$ eV ($\Delta H_f =$ 1.48 eV, Fig. S9(c)), indicating Fermi level pinning that stabilizes $H_{i(AB)}^{+1}$ and $H_{i(AB)}^{-1}$ over the neutral state ($\Delta H_f = 1.86$ eV). However, the high migration barriers—1.44 eV for $H_{i(AB)}^{+1}$ (Fig. S12(g)) and 1.17 eV for $H_{i(AB)}^{-1}$ (Fig. S12(h))—are 20–50% higher than those at the BC site (0.96–1.06 eV), effectively trapping hydrogen at AB configurations. This kinetic constraint, combined with the elevated formation energies, indicates that AB sites act as deep traps rather than diffusion pathways.

At the M site, hydrogen favors the negatively charged state $H_{i(M)}^{-1}$, with a formation energy (Fig. S9(e)) below 1.5 eV—lower than $H_{i(M)}^{0}$ (1.72 eV) and $H_{i(M)}^{+1}$ (>2 eV). However, the migration barrier for $H_{i(M)}^{-1}$ is 1.37 eV (Fig. S13(c)), 43% higher than that of $H_{i(BC)}^{0}$ (0.96 eV), while $H_{i(M)}^{+1}$ exhibits a prohibitive barrier of 4.64 eV (Fig. S12(e)). Despite favorable thermodynamics, M-site hydrogen is therefore kinetically trapped and cannot contribute to long-range diffusion.

The T and C sites exhibit similar energetics (Fig. S9(a) and Fig. S9(f)), with $H_{i}^{-1}$ having formation energies below 1.5 eV—lower than $H_i^0$ (1.68 eV for T, 1.72 eV for C). Migration barriers for $H_{i(T)}^{-1}$ (1.03 eV, Fig. S13(d)) and $H_{i(C)}^{-1}$ (1.08 eV, Fig. S13(b)) are 7–13% higher than $H_{i(BC)}^0$ (0.96 eV), while $H_{i(T)}^{+1}$ (1.22 eV, Fig. S12(c)) and $H_{i(C)}^{+1}$ (1.06 eV, Fig. S12(f)) show similar elevations. Notably, $H_{i(T)}^0$ exhibits an anomalously low migration barrier of 0.71 eV (Fig. S12(a))—26% lower than $H_{i(BC)}^0$—suggesting facile neutral hydrogen diffusion at T sites. However, the substantially higher formation energy (1.68 eV vs. 1.42 eV at BC) renders $H_{i(T)}^0$ thermodynamically unfavorable, with negligible equilibrium populations. This kinetic-thermodynamic mismatch—where low barriers coexist with high formation energies—prevents T and C sites from contributing significantly to hydrogen transport.

Comparison across all interstitial sites reveals that the BC configuration uniquely balances thermodynamic stability with kinetic accessibility. While alternative sites (M, T, C) exhibit competitive formation energies (<1.5 eV for charged states), their migration barriers are 7–43% higher than $H_{i(BC)}^0$, ranging from 1.03 eV to 1.37 eV. The AB site exhibits even more restrictive energetics (1.48–1.86 eV) and migration barriers (1.17–1.44 eV). Although $H_{i(T)}^0$ has a lower barrier (0.71 eV), its higher formation energy (1.68 eV vs. 1.42 eV at BC) results in negligible populations.

The BC site's moderate formation energies (~1.42 eV) and lowest migration barriers (0.96–1.06 eV), combined with mid-gap Fermi level pinning of the neutral state, establish $H_{i(BC)}^0$ as the dominant species governing hydrogen transport in intrinsic silicon. The following sections examine its interactions with native defects.

**Equilibrium Concentration Analysis for Hydrogen Interstitials.** While the diffusion analysis above establishes the kinetic pathways for hydrogen migration, it does not directly indicate the actual defect populations within the crystal under equilibrium conditions. To provide a more complete assessment, we calculated the equilibrium concentrations of $H_i$ defects in c-Si at Fermi level $E_f = 0$. The total degeneracy $g$ for each charge state was determined using Eq. (13), with values detailed in Table 4. The $H_{i(BC)}^{\pm 1}$ sites exhibit $C_S$ point group symmetry, whereas the $H_{i(BC)}^0$ possesses $D_{3d}$ symmetry.

**Table 4** Degeneracy factors $g$ for $H_{i(BC)}$ interstitial at different charge states in silicon.

| | Charge states | | |
|---|---|---|---|
| | +1 | 0 | -1 |
| $g_{sites}$ | 24 | 4 | 24 |
| $g_{orient}$ | 12 | 2 | 12 |
| $g_{irrep}$ | 2 | 8 | 2 |
| $g_{spin}$ | 1 | 1 | 1 |
| $g_{total}$ | 576 | 64 | 576 |

The equilibrium concentrations calculated using Eq. (12) are shown in Fig. S10(a). $H_{i(BC)}^0$ exhibits the highest formation energy (1.42 eV at $E_f = 0$) combined with the lowest degeneracy (64 vs. 576), resulting in significantly lower concentration compared to the charged states. In contrast, $H_{i(BC)}^{\pm 1}$ benefit from both lower formation energies and higher degeneracy (576), which provides more equivalent configurations for stable defect existence, leading to higher equilibrium concentrations. All concentrations increase with temperature, facilitating $H_i$ incorporation into c-Si.

Our calculations yield hydrogen concentrations of $10^{16}\ \mathrm{cm}^{-3}$ at 433 K, substantially lower than experimental values[72] ($10^{17}\ \mathrm{cm}^{-3}$ at 433 K). This discrepancy reflects the limitation of our BC-site-specific model compared to experimental measurements that encompass hydrogen distributed across all possible lattice and interstitial positions. The relatively low diffusion barrier of $\mathrm{H}_{i(\mathrm{BC})}^{0}$ (0.9 eV) enables hydrogen atoms to readily migrate within the Si lattice, facilitating diffusion from hydrogen-rich environments (such as aqueous solutions) into the Si interior and leading to the observed high experimental concentrations.

**Electronic Structure and Deep-Level Defect States.** Beyond equilibrium concentrations, understanding the electronic properties of hydrogen defects is essential for evaluating their impact on silicon device performance. Fig. 2(d) shows the electronic states of $\mathrm{H}_{i(\mathrm{BC})}^{0}$ near the Fermi level, revealing contributions from Si 3s/3p and H 1s orbitals. The hydrogen defect introduces a deep level located 0.24 eV below the conduction band minimum (CBM), classifying it as a carrier recombination center. Such deep-level defects trap charge carriers and promote non-radiative recombination, directly reducing carrier lifetimes and degrading the performance of silicon-based devices such as solar cells and transistors. [73]

Electronic structure calculations for hydrogen at other interstitial sites (Fig. S14) reveal that hydrogen defects consistently introduce detrimental electronic states—either deep levels or resonance states—regardless of their lattice position. This universality demonstrates that hydrogen incorporation fundamentally compromises silicon's electronic properties across all investigated sites, providing comprehensive theoretical validation for the observed performance degradation in hydrogen-exposed silicon devices.

**Non-Radiative Recombination and Photovoltaic Performance.** To understand the carrier recombination behavior of $\mathrm{H}_{i(\mathrm{BC})}$, we performed non-radiative recombination calculations using the theoretical framework described above. Fig. 3 presents a comprehensive analysis of the recombination properties through four complementary perspectives.

The configuration coordinate diagram (CCD) (Fig. 3(a)) reveals that $\mathrm{H}_{i(\mathrm{BC})}$ exhibit relatively weak electron-phonon coupling characteristics. The temperature-dependent capture coefficients (Fig. 3(b)) span broad ranges over 250-650 K, with electron capture coefficients from $1.91 \times 10^{-10}$ to $2.47 \times 10^{-9}$ and hole capture coefficients from $2.71 \times 10^{-15}$ to $3.49 \times 10^{-14}\,\mathrm{cm}^{3}/\mathrm{s}$. The effective capture coefficients all in the moderate recombination regime. The corresponding capture cross sections (Fig. 3(c)) show electron values between $8.45 \times 10^{-18}$ and $6.79 \times 10^{-17}\mathrm{cm}^{2}$, and hole cross sections from $1.20 \times 10^{-22}$ to $9.64 \times 10^{-21}\mathrm{cm}^{2}$. These effective cross sections indicate moderate electron trapping and very weak hole trapping capabilities. The net SRH recombination coefficient $R_{\mathrm{SRH}}$ (Fig. 3(d)) spans from $1.34 \times 10^{8}$ to $8.84 \times 10^{21}\mathrm{cm}^{-3}/\mathrm{s}$. The recombination rates are consistent with the moderate effective capture coefficients observed in Fig. 3(b). The calculated moderate recombination coefficients confirm that $\mathrm{H}_{i(\mathrm{BC})}$ indeed act as effective recombination centers, consistent with the deep energy levels identified in the electronic structure analysis.

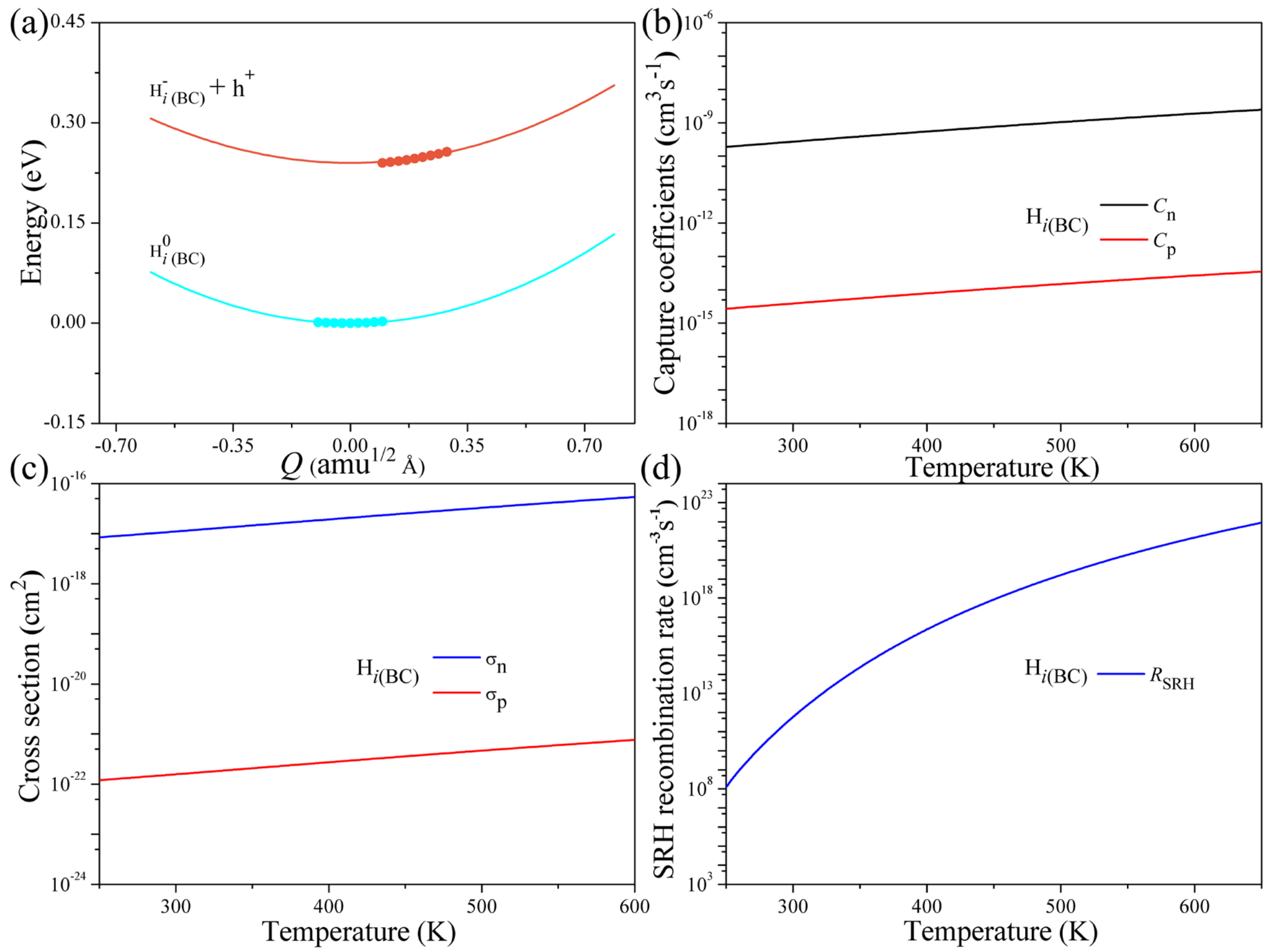

**Figure 3**. Non-radiative recombination properties of $H_{i(BC)}$: (a) configuration coordinate diagram (CCD) for hydrogen interstitial defects, (b) capture coefficients for electrons ($C_n$) and holes ($C_p$), (c) capture cross sections for electrons ($\sigma_n$) and holes ($\sigma_p$), and (d) net SRH recombination rate $R_{SRH}$.

Having established that hydrogen defects at BC sites exhibit moderate electron trapping and weak hole trapping capabilities, we now investigate their impact on photovoltaic device performance through transport theory calculations. The photovoltaic current of hydrogen-doped silicon-based solar cells was calculated. The introduction of hydrogen defects into c-Si significantly impacts its photovoltaic properties, as shown in Fig. 2(e). To establish an appropriately sized device model, the concentration of hydrogen interstitial defects in the device was set to $7.64 \times 10^{20} cm^{-3}$, as determined from the geometry of the simulated scattering region (see Section S4 for details). This concentration is employed as a model parameter in the quantum transport calculations. Our calculations, based on quantum transport theory, reveal that hydrogen doping, regardless of the specific lattice site, leads to a significant reduction in photocurrent intensity. From Fig. 2(d) and Fig. 2(e), it is evident that the defect states introduced by hydrogen are deep-level defects and carrier recombination centers, which negatively affect the photoelectric performance of c-Si, substantially decreasing the photocurrent intensity. The calculated non-radiative recombination capture coefficients confirm this behavior, showing moderate electron capture rates (from $1.91 \times 10^{-10}$ to $2.47 \times 10^{-9} cm^3/s$) and substantial SRH recombination activity that validate the deep-level nature of these defects. This decline suggests that hydrogen defects act as effective recombination centers that hinder the efficient transport of photogenerated charge carriers, which is crucial for optimal photovoltaic performance.

Our optical absorption coefficient calculations reveal that $H^0_{i(BC)}$-containing Si exhibits only slightly increases compared to pure silicon, as shown in Fig. S16(a)(b), indicating that enhanced optical absorption is not the primary cause of the photocurrent reduction. Instead, the detrimental role of hydrogen defects primarily disrupts charge carrier mobility and recombination processes through non-radiative pathways, ultimately compromising the overall efficiency of c-Si in photovoltaic applications. This mechanism explains one of the reasons why water reduces the efficiency of silicon-based solar cells.

## 3.2 Impact of O interstitial defects

Previous studies have shown that the occupancy site of interstitial oxygen in c-Si differs from that of hydrogen, as oxygen occupies only the BC site within Si.[24, 25] There are six equivalent BC sites for oxygen in c-Si, all considered equivalent. Additionally, previous research[15] indicates that during diffusion, oxygen prefers the BC1 site configuration. Therefore, the initial and final states of diffusion can be studied at the BC1 site. The distance between O and the nearest Si atoms is 1.563 Å, and the Si-O-Si bond angle is 98°.

**Formation Energy Analysis for Oxygen Defects.** To characterize oxygen interstitial defects analogous to our hydrogen analysis, we employ the same theoretical methodology with appropriate chemical potential constraints. The thermodynamic stability range must first be established through chemical potential analysis. The oxygen chemical potential range is constrained to prevent $O_2$ precipitation and Si-oxide phase formation (SiO, $SiO_2$, $SiO_4$), defining the accessible region for $\Delta\mu_{\mathrm{O}}$ (oxygen chemical potential) and $\Delta\mu_{\mathrm{Si}}$ (silicon chemical potential) as illustrated in Fig. 4(a). This thermodynamic framework ensures that only isolated oxygen interstitials are considered, excluding competing phases that could influence defect formation energies.

Using the oxygen-rich chemical potential (-9.14 eV) in Eq. (1), we calculated the formation energies for oxygen interstitials in four charge states (-2, -1, 0, +1), as shown in Fig. 4(b). The results reveal that $O^{-2}_{i(BC1)}$, $O^{-1}_{i(BC1)}$, and $O^{0}_{i(BC1)}$ are the thermodynamically favorable states for BC1 site occupancy, consistent with experimental evidence that individual oxygen atoms preferentially occupy interstitial bond-center (BC) positions in silicon[74]. Our calculated formation energy for neutral oxygen at the BC site is 1.60 eV, in good agreement with the 1.66 eV value reported by Coutinho *et al.* using density functional theory with similar computational parameters[75, 76].

As shown in Fig. 4(b), the charge state stability varies systematically with Fermi level position: $O^{0}_{i(BC1)}$ dominates when $E_f$ ranges from -0.25 to 0.21 eV, $O^{-1}_{i(BC1)}$ becomes most stable between 0.21 and 0.38 eV, and $O^{-2}_{i(BC1)}$ prevails when $E_f$ exceeds 0.38 eV. The calculated charge transition levels are $\varepsilon(0/-1) = 0.21$ eV and $\varepsilon(-1/-2) = 0.38$ eV according to Eq. (14). The multiple negative charge states and their relatively low transition levels suggest that oxygen interstitials exhibit strong acceptor-like behavior, capable of trapping electrons and influencing Fermi level positioning in silicon.

$\mathrm{O}_{i(\mathrm{BC1})}$ in silicon behaves as an acceptor-like defect that can trap electrons when the Fermi level rises. As the Fermi level increases, $\mathrm{O}_{i(\mathrm{BC1})}$ tends to capture electrons and transform into negatively charged states ($O^{-1}_{i(BC1)}$ or $O^{-2}_{i(BC1)}$), thereby pulling the Fermi level back toward the lower part of the band gap. This electron-trapping behavior effectively stabilizes a relatively low Fermi-level position, indicating that $\mathrm{O}_{i(\mathrm{BC1})}$ compensates donor impurities and suppresses n-type conductivity while slightly promoting p-type characteristics. Consequently, the Fermi level is likely to be constrained within the range of -0.25 to 0.21 eV, which increases the likelihood of the formation of $O^{0}_{i(\mathrm{BC1})}$low Fermi-level position, indicating that $O_{i(\mathrm{BC1})}$ compensates donor impurities and suppresses n-type conductivity while slightly promoting p-type characteristics. Consequently, the Fermi level is likely to be constrained within the range of -0.25 to 0.21eV, which increases the likelihood of the formation of $O^{0}_{i(\mathrm{BC1})}$.

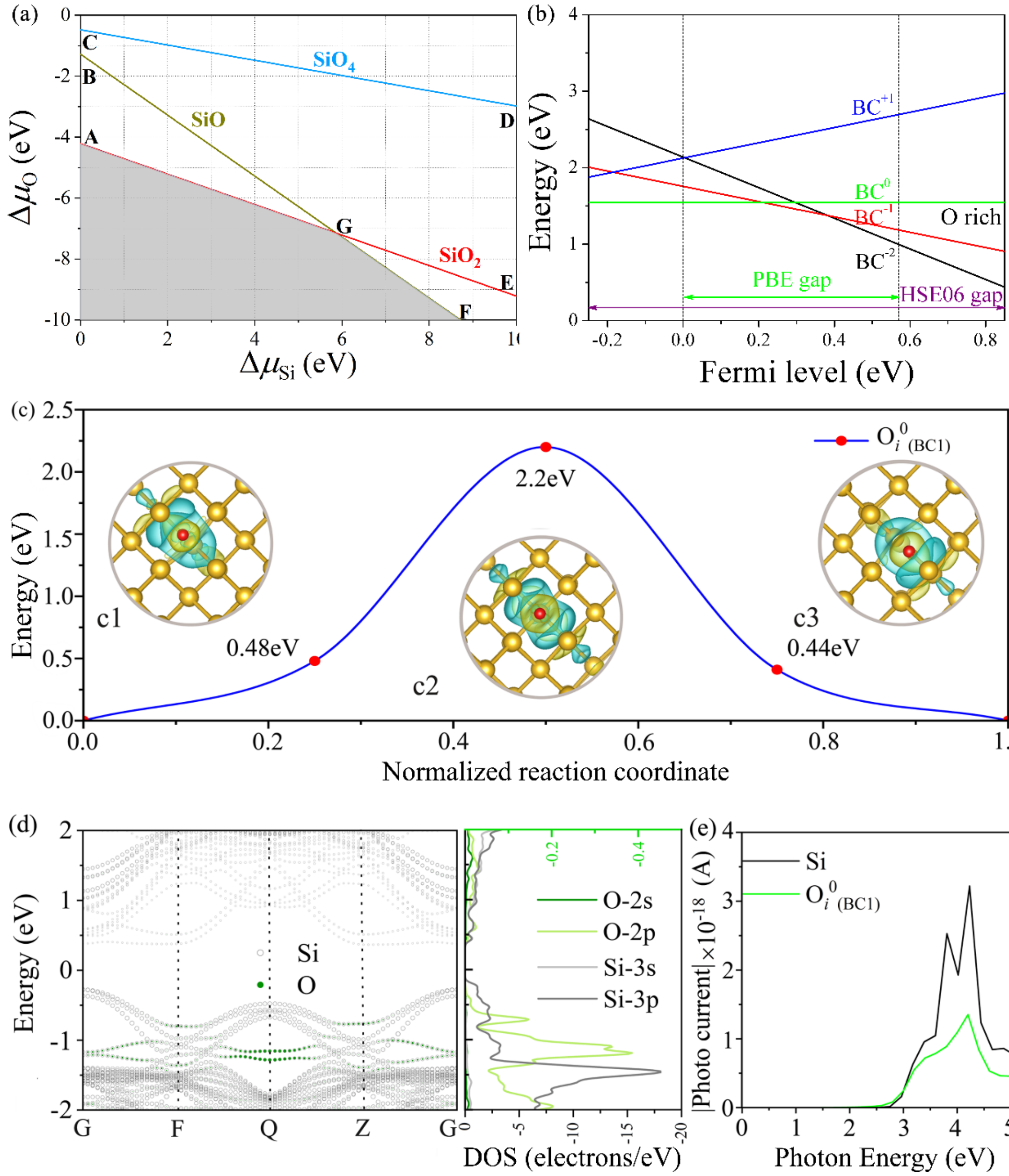

**Figure 4**. The chemical potential range for (a) $\Delta\mu_{\mathrm{Si}}$ and $\Delta\mu_{\mathrm{O}}$ doping in the c-Si doping in the c-Si; (b) the defect formation energies of $O^{-2}_{i(\mathrm{BC1})}$, $O^{-1}_{i(\mathrm{BC1})}$, $O^{0}_{i(\mathrm{BC1})}$, $O^{+1}_{i(\mathrm{BC1})}$, $O^{+2}_{i(\mathrm{BC1})}$; (c) the diffusion barrier of the pathway for oxygen from the BC1 site to the nearest BC1 site, along with the corresponding differential charge density for the diffusion path; and (d) the combination of projected bands and projected density of states for $O^{0}_{i(\mathrm{BC1})}$; (e) photocurrent intensity figure of neutral hydrogen dopant

defects at different sites with a device cross-sectional area of $1.20 \times 10^{-18}$ m$^2$.

Fig. 4(a) establishes the thermodynamic stability window for oxygen defects by constraining the chemical potential to prevent oxide precipitation: $-\infty < \Delta\mu_{\mathrm{O}} < -4.22$ eV. The AGFOA region represents the phase-pure domain where only isolated oxygen interstitials exist without competing secondary phases. According to Eq. (3), the oxygen-rich limit corresponds to point A with $\mu_{\mathrm{O}} = -9.14$ eV, which serves as the reference condition for formation energy calculations.

**Oxygen Interstitial Migration and Kinetic Constraints**. Using the CINEB method, we calculated the diffusion barriers and pathways for oxygen interstitials in the 0, -1, and -2 charge states between BC1 sites. The diffusion pathway exhibits asymmetric geometry, consistent with previous literature findings[15]. The results reveal a systematic reduction in diffusion barriers with increasing negative charge: $\mathrm{O}_{i(\mathrm{BC1})}^{0}$ exhibits a substantial barrier of 2.2 eV (Fig. 4(c)), which decreases to 2.02 eV for $\mathrm{O}_{i(\mathrm{BC1})}^{-1}$ (Fig. S15(c)) and further to 1.74 eV for $\mathrm{O}_{i(\mathrm{BC1})}^{-2}$ (Fig. S15(d)). This trend reflects enhanced mobility as oxygen captures additional electrons, reducing the repulsive interactions with surrounding silicon atoms during migration[77].

However, the Fermi level pinning effect discussed above significantly influences the practical diffusion scenario. The charge state stability transitions occur at $\varepsilon(0/-1) = 0.21$ eV and $\varepsilon(-1/-2) = 0.38$ eV, with corresponding formation energies decreasing from 1.56 eV $\mathrm{O}_{i(\mathrm{BC1})}^{0}$ at $E_f < 0.21$ eV) to 1.38 eV ($\mathrm{O}_{i(\mathrm{BC1})}^{-1}$ at $0.21 - 0.38$ eV) and below 1.38 eV for $\mathrm{O}_{i(\mathrm{BC1})}^{-2}$ ($E_f > 0.38$ eV). Under typical near-intrinsic conditions, the acceptor-like behavior of oxygen interstitials stabilizes the Fermi level in the lower band gap region ($E_f = -0.25 \sim 0.21$ eV), making $\mathrm{O}_{i(\mathrm{BC1})}^{0}$ the predominant charge state despite its highest diffusion barrier. While negatively charged states exhibit lower migration barriers (2.02 eV and 1.74 eV), their limited equilibrium concentrations at low Fermi levels prevent them from dominating oxygen transport. Consequently, oxygen diffusion in silicon is primarily governed by the slow migration of neutral interstitials with the 2.2 eV barrier, explaining the relatively low oxygen incorporation observed experimentally in aqueous environments compared to neutral hydrogen[24]. This high diffusion barrier effectively limits oxygen penetration into silicon, contrasting sharply with the facile hydrogen diffusion (0.96 eV barrier) discussed previously.

**Equilibrium Concentration Analysis for Oxygen Interstitials.** Beyond these kinetic constraints, we calculated the equilibrium concentrations of $\mathrm{O}_{i(\mathrm{BC1})}$ as a function of temperature to assess the actual defect populations achievable in silicon. The total degeneracy $g$ for all charge states (0, -1, -2) was determined as 576 using Eq. (13). The equilibrium concentrations calculated using Eq. (12) show that $\mathrm{O}_{i(\mathrm{BC1})}^{-2}$ exhibits the highest concentration, followed by $\mathrm{O}_{i(\mathrm{BC1})}^{-1}$ and $\mathrm{O}_{i(\mathrm{BC1})}^{0}$. This trend directly correlates with the formation energy hierarchy: as the Fermi level increases, the formation energies follow $\Delta H_f[\mathrm{O}_{i(\mathrm{BC1})}^{0}] > \Delta H_f[\mathrm{O}_{i(\mathrm{BC1})}^{-1}] > \Delta H_f[\mathrm{O}_{i(\mathrm{BC1})}^{-2}]$, making more negatively charged states energetically favorable. Since all charge states share identical degeneracy, the concentration differences are primarily governed by formation energy variations.

Our calculations yield oxygen concentrations on the order of $10^{15}$ cm$^{-3}$ at 300 K (Fig. S15(a)), approaching the experimental range of $10^{17}$ cm$^{-3}$ observed during Si crystal growth,[78] demonstrating good agreement between theoretical predictions and experimental observations. However, despite the favorable equilibrium concentrations, oxygen incorporation faces significant kinetic limitations under ambient conditions due to both the high diffusion barrier of $O^0_{i(BC1)}$ (2.2 eV) and Fermi level pinning that restricts formation of the more mobile charged states. This kinetic bottleneck explains why oxygen defects require high-temperature growth conditions for substantial incorporation, contrasting with the more facile hydrogen diffusion processes. Consequently, oxygen from aqueous environments cannot readily penetrate silicon under normal processing conditions, with significant oxygen concentrations achievable only during high-temperature crystal growth when kinetic barriers can be overcome.

**Electronic Structure Analysis of Oxygen Defects**. Due to their thermodynamic stability, neutral oxygen atoms ($O^0_{i(BC1)}$) represent the dominant defect species in silicon. Electronic structure analysis reveals that oxygen defects at BC1 sites create resonant states at the valence band edge (Fig. 4d), contrasting with the deep-level trap states of hydrogen interstitials. These resonant states, strongly coupled to the valence band, facilitate carrier scattering and non-radiative recombination through enhanced electron-phonon interactions.[79] However, their recombination activity is weaker than deep-level hydrogen defects due to their band-edge position rather than mid-gap placement. Quantitative assessment of these carrier capture mechanisms requires detailed analysis of the recombination parameters.

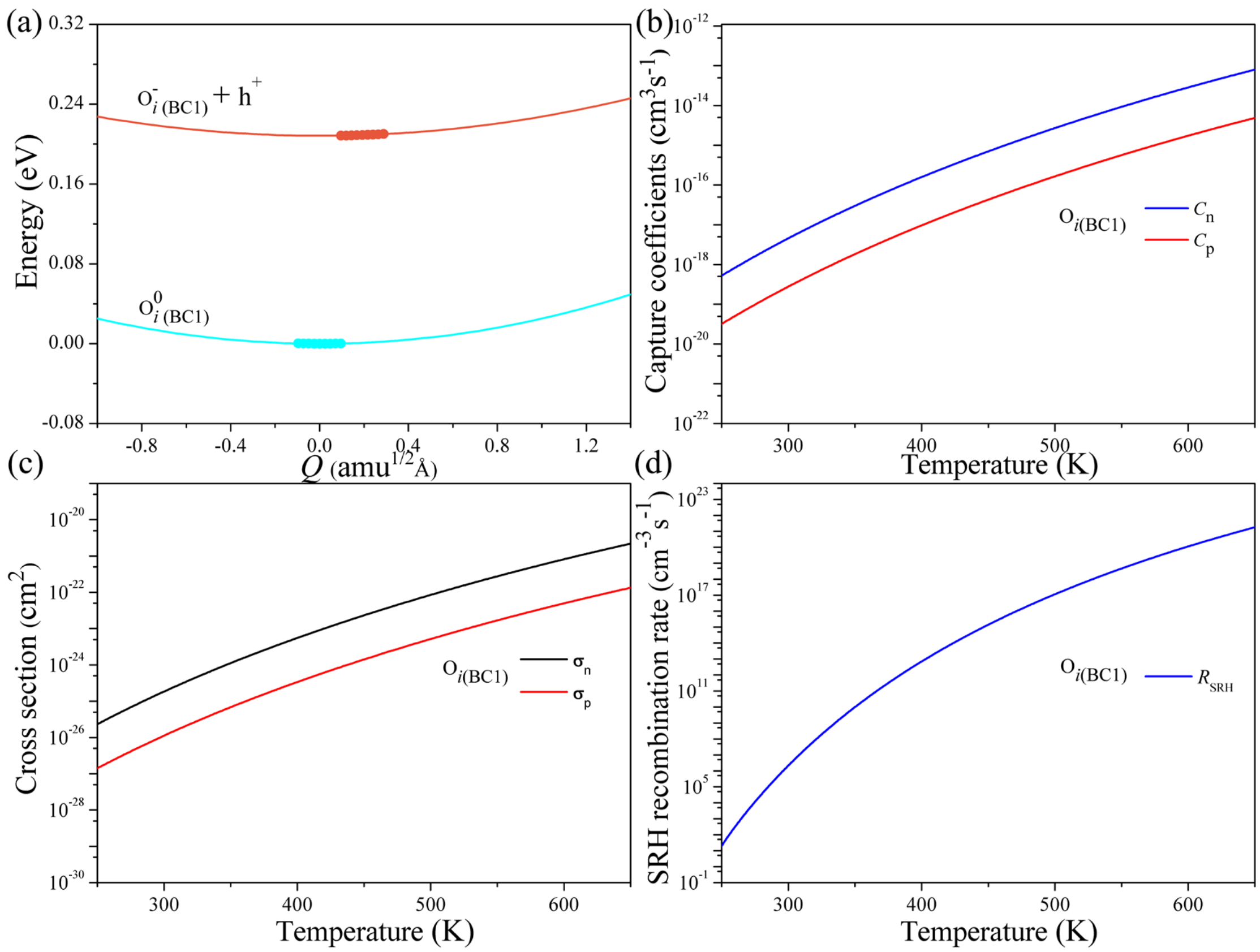

**Figure 5**. Non-radiative recombination properties of $\mathrm{O}_{i(\mathrm{BC1})}$: (a) configuration coordinate diagram (CCD) for oxygen interstitial defects, (b) capture coefficients for electrons ($C_\mathrm{n}$) and holes ($C_\mathrm{p}$), (c) capture cross sections for electrons ($\sigma_\mathrm{n}$) and holes ($\sigma_\mathrm{p}$), and (d) net SRH recombination rate $R_\mathrm{SRH}$.

**Non-Radiative Recombination and Photovoltaic Performance Analysis.** To further investigate the impact of oxygen defects on carrier recombination, we performed similar non-radiative recombination calculations for $O_{i(BC1)}$ (Fig. 5). The CCD in Fig. 5(a) reveals that $O_{i(BC1)}$ exhibit weak electron-phonon coupling, indicating relatively small structural relaxation upon carrier capture. This weak coupling behavior contrasts with the moderate coupling observed for $H_{i(BC)}$. The calculated non-radiative capture coefficients shown in Fig. 5(b) demonstrate that $O_{i(BC1)}$ exhibit significantly weaker recombination activity compared to hydrogen defects. The electron capture coefficients range from $5.34 \times 10^{-19}$ to $7.97 \times 10^{-14}\,\mathrm{cm^3/s}$, while hole capture coefficients span from $3.26 \times 10^{-20}$ to $4.86 \times 10^{-15}\,\mathrm{cm^3/s}$ across the temperature range. The corresponding capture cross-sections in Fig. 5(c) further confirm this weak recombination behavior, with electron capture cross-sections ranging from $2.36 \times 10^{-26}$ to $2.23 \times 10^{-21}\mathrm{cm^2}$ and hole capture cross-sections from $1.44 \times 10^{-27}$ to $1.36 \times 10^{-22}\mathrm{cm^2}$. These values are substantially smaller than those observed for hydrogen defects, indicating less effective carrier trapping. The total SRH recombination rate $R_{\mathrm{SRH}}$ shown in Fig. 5(d) ranges from $2.2 \times 10^{1}$ to $1.79 \times 10^{21}\mathrm{cm^{-3}/s}$, which, despite the wide range, represents significantly lower recombination activity compared to hydrogen defects at similar defect concentrations.

The weak electron-phonon coupling and correspondingly low capture coefficients confirm that $O_{i(BC1)}$ act as relatively benign defects with minimal impact on carrier recombination. These non-radiative recombination results are fully consistent with the above electronic structure calculations, which identified oxygen defects as resonant states and weak recombination centers. The calculated capture coefficients, being several orders of magnitude lower than those of hydrogen defects, quantitatively validate the electronic structure prediction that oxygen defects introduce only shallow or resonant levels with limited carrier trapping capability. Our optical absorption coefficient calculations further support this conclusion, showing that $O^{0}_{i(BC1)}$-containing Si exhibits negligible changes compared to pure silicon, as demonstrated in Fig. S16(b)(c). This indicates that altered optical absorption is not responsible for the observed photocurrent reduction in oxygen-defected Si. Instead, the modest photocurrent decline is primarily attributed to the limited carrier trapping effects of oxygen's shallow/resonant states. This agreement between electronic structure, optical properties, and recombination dynamics consistently confirms that oxygen defects are indeed less detrimental to photovoltaic performance compared to the deep-level hydrogen defects.

This effect is particularly detrimental in devices where high carrier mobility is crucial, such as solar cells, as it leads to non-radiative recombination pathways that reduce overall device efficiency. Additionally, these defect-induced trapping mechanisms can increase recombination rates, decreasing photocurrent in optoelectronic applications, as shown in Fig. 4(e).[80] Our device simulations reveal that when $O_{i(BC1)}$ are present at high concentrations in Si, the photovoltaic current decreases compared to pure silicon, though the reduction is less pronounced than that observed for $H_{i(BC)}$ defects under similar conditions. This relatively limited impact suggests that while oxygen-related resonant states do degrade device performance, their detrimental effects are somewhat less severe than those of hydrogen defects. Importantly, the significant kinetic barriers limiting oxygen diffusion at ambient temperatures mean that oxygen from aqueous environments has minimal impact on Si solar cell performance under normal operating conditions. Unlike hydrogen atoms which readily diffuse into Si at room temperature, oxygen incorporation requires high-temperature crystal growth environments to achieve substantial concentrations. Consequently, the influence of waterborne oxygen on silicon solar cell efficiency is negligible under typical ambient conditions, making oxygen contamination a much less critical concern than hydrogen-related degradation in photovoltaic applications.

# 4. Conclusion

We have systematically investigated the microscopic origins of MID effects in silicon solar cells through comprehensive first-principles analysis of hydrogen and oxygen interstitial defects from aqueous environments. Our theoretical framework elucidates the distinct roles of waterborne H and O species in photovoltaic performance degradation.

For hydrogen-induced MID effects, $\mathrm{H}_{i(\mathrm{BC})}^{0}$ defects readily incorporate from moisture due to their moderate diffusion barrier (0.96 eV for $\mathrm{H}_{i(\mathrm{BC})}^{0}$), enabling significant hydrogen penetration at processing temperatures. The formation of deep-level mid-gap states creates efficient non-radiative recombination centers with calculated capture coefficients of $10^{-15}$ to $10^{-9}\mathrm{cm}^3/\mathrm{s}$ and SRH recombination rates of $10^8$ to $10^{21}$ $\mathrm{cm}^3/\mathrm{s}$. These deep-level traps substantially enhance minority carrier recombination, directly causing the photocurrent degradation characteristic of hydrogen-induced MID in silicon solar cells.

In contrast, oxygen-induced MID effects are severely limited by kinetic constraints. Despite thermodynamically favorable formation with calculated equilibrium concentrations (~$10^{16}$ $\mathrm{cm}^{-3}$) matching experimental values (~$10^{17}\,\mathrm{cm}^{-3}$), the high diffusion barrier (2.2 eV) of $\mathrm{O}_{i(\mathrm{BC1})}^{0}$ prevents substantial oxygen incorporation from moisture under normal processing conditions. Furthermore, oxygen defects create resonant states at the valence band edge rather than deep-level traps, resulting in weak capture coefficients ($10^{-20}$ to $10^{-14}$ $\mathrm{cm}^3/\mathrm{s}$) that are several orders of magnitude lower than hydrogen defects. Consequently, oxygen contributes minimally to MID effects in moisture-exposed silicon solar cells.

These findings establish hydrogen interstitials as the dominant pathway for moisture-induced MID while confirming the negligible role of oxygen under typical operating conditions. The stark contrast between hydrogen's facile incorporation and strong recombination activity versus oxygen's kinetic limitations and weak recombination behavior explains the selective nature of moisture-induced MID effects. Our multi phonon recombination analysis quantitatively validates that hydrogen-induced MID originates from deep-level trap formation, while kinetically limited oxygen incorporation prevents significant oxygen-related degradation.

The theoretical framework provides essential guidance for mitigating MID in silicon solar cell manufacturing. Control strategies should focus primarily on minimizing hydrogen exposure from moisture environments, while oxygen contamination poses minimal risk under normal processing temperatures. These insights enable targeted optimization of silicon photovoltaic processing protocols to minimize MID effects and maintain long-term device performance.

**Supporting Information**

Details for Makov-Payne correction method; Details for Band energy corrections method; CINEB computational details; Device modeling computational details (PDF)

**Acknowledgements**

This research was supported by National Natural Science Foundation of China (No. 12074096 and No. 11747103), Science and Technology Project of Hebei Education Department (No. QN2022149) and Key Project of Natural Science Foundation of Hebei Province (B2025208103).